\documentclass[11pt,twoside]{article}
\pdfoutput=1
\usepackage{epsfig}
\usepackage{rotate}
\usepackage[figuresright]{rotating}
\usepackage{lscape}
\usepackage{hyperref}
\usepackage{url}
\usepackage{amssymb}
\usepackage{graphicx}
\usepackage{cite}

\usepackage{multirow}
\usepackage{capt-of}
\usepackage{wrapfig}            %Wrapping figures
\usepackage{wasysym}            %for the \diameter symbol
\usepackage{textcomp}
\DeclareFontFamily{OML}{eur}{\skewchar\font127} \DeclareFontShape{OML}{eur}{m}{n}{<5> <6> 
  <7> <8> <9> gen * eurm <10><10.95><12><14.4><17.28><20.74><24.88>eurm10}{} 
\DeclareSymbolFont{greek}{OML}{eur}{m}{n} 
\DeclareMathSymbol{\upmu}{\mathord}{greek}{"16}
\begin{document}
\title{\textsc{A measurement of $\sigma_{Zh}$ at a future $e^+e^-$ collider using the hadronic decay of $Z$}} %% 

%***********************************************************************
% AUTHORS INFORMATION AREA
%***********************************************************************
\author{Akiya Miyamoto\footnote{e-mail: akiya.miyamoto@kek.jp}
\vspace{.3cm}\\
High Energy Accelerator Research Organization (KEK),\\
1-1 Oho, Tsukuba, Ibaraki, 305-0801 Japan
}
\date{}
%***********************************************************************
% END OF AUTHORS INFORMATION AREA
%***********************************************************************

\maketitle

\begin{abstract}
A feasibility to use the hadronic decay mode of $Z$ for the model independant measurement of 
the total cross section of Higgs-strahlung process $(\sigma_{Zh})$ at a future $e^+e^-$ collider was studied.
For the recoil mass measurement from hadronic decay of $Z$, a simple cut based analysis was applied 
on samples produced by the ILD full detector simulation
at $\sqrt{s}=350$ GeV and 500 GeV using the ILC beam parameters.
At $350$ GeV, a bump in the recoil mass distribution was reconstructed, and 
$\Delta\sigma_{Zh}/\sigma_{Zh}$ = 3.4\% was obtained 
assuming 165 fb$^{-1}$ data with $e^-(e^+)$ beam polarization 
of -80\%(+30\%) and +80\%(-30\%), respectively. 
At $500$ GeV, clear Higgs boson peak in the recoil mass distribution was not seen, 
however, from the excess of the events, $\Delta\sigma_{Zh}/\sigma_{Zh}=3.9\%$ was obtained
assuming 500 fb$^{-1}$ data with  $e^-(e^+)$ beam polarization of -80\%(+30\%).
\end{abstract}

%%%%%%%%%%%%%%%%%%%%%%%%%%%%%%%%%%%%%%%%%%%%%%%%%%%%%%%%%%%%%%%%%%%%%%%%%%%
\section{Introduction}
The total cross section of the Higgs-strahlung 
process ($\sigma_{Zh})$ can be measured model independently 
using the recoil mass technique at a future $e^+e^-$ collider.
The $ZZh$ coupling strength derived from this measurement   
is a crucial input for investigating physics beyond the standard model
though Higgs properties.  

Previously, the recoil mass of the Higgs-strahlung process were studied for the case 
where $Z$ decays to $e^+e^-$ or $\mu^+\mu^-$ pairs; $Z$ momentum can be measured 
precisely thanks to a high precision tracking detector 
and a good signal to noise ratio in the recoil mass measurement has been reported ( see ref.\cite{Baer:2013cma} 
and reference therein ).
In the case of ILC, 2.6\% of the measurement precision 
for the total cross section, $\Delta\sigma_{Zh} / \sigma_{Zh}$, was 
expected from the recoil mass measurement by combining $e^+e^-$ and $\mu^+\mu^-$ 
channel at $\sqrt{s}=250$ GeV with 250~fb$^{-1}$ data\cite{Baer:2013cma,higgs-LC-white-paper}.
At 350 GeV, the recoil mass resolution is worse than the 250~GeV case, but
similar precision\footnote{
About 10\% worse precision would be expected  
because the result on $\mu^+\mu^-h$ in ref.\cite{HengneLi:2010} could be scaled by ILC TDR luminosity to 
$\Delta\sigma_{Zh}/\sigma_{Zh}$ of 3.4\% for 250 fb$^{-1}$ at 250 GeV 
and 3.7\% 330 fb$^{-1}$ at 350 GeV, respectively. 
}
of the total cross section measurement is expected from a fast simulation study of $\mu^+\mu^-h$ 
channel\cite{HengneLi:2010} thanks to the better signal selection efficiency and the higher luminosity 
provided by ILC.
At 500 GeV, the performance is degraded by further worse recoil mass resolution and increased standard model 
background, still a preliminary result of $\Delta\sigma_{Zh}/\sigma_{Zh}$=4.8\% combining $e^{+}e^{-}$ and 
$\mu^{+}\mu^{-}$ channels were reported assuming 500 fb$^{-1}$ data with -80\%(+30\%) $e^{-}(e^{+})$ beam
polarization\cite{Taikan-llh-500,higgs-LC-white-paper}.
The recoil mass measurement by using the leptonic decay mode of $Z$ is limited by the 
small branching ratio of $Z$. In order to achieve a precision close to 1\%, 
a measurement with higher luminosity has been proposed\cite{higgs-LC-white-paper}.

The branching ratio of $Z$ to quark pair is about factor 10 larger than the sum of $e^+e^-$ and $\mu^+\mu^-$  mode.
In this paper, we study a feasibility of recoil mass measurement using hadronic decay mode of $Z$.
The detectors for ILC are equipped with a particle flow calorimeter, aiming 
to achieve the jet energy resolution($\Delta E/E$) better than $30\%/\sqrt{E({\rm GeV})}$.
The high precision jet energy measurement is crucial for this study.  

Note that the recoil mass of Higgs-strahlung process is given by 
$m_{h}^2=E_{cm}^2 - 2E_{cm}E_{Z} + m_{Z}^2$ when the effect of beamstrahlung and bremsstrahlung is neglected,
where $m_h$, $m_Z$, $E_{cm}$, and $E_{Z}$ are the mass of higgs, $Z$, the center of mass energy and 
the energy of $Z$.  Therefore, the recoil mass resolution, $\Delta m_{h}$, is given by 
$\Delta m_{h} = { ( E_{cm} /  m_{h} ) } \Delta E_{Z}$.
If $E_{Z}$ is measured by a PFO calorimeter, the relative energy resolution of jet is 
almost independent of the jet energy; in the case of ILD, $\Delta E_{Z}/E_{Z} \sim 3\%$ for $E_{Z}$ 
from 90 to 500 GeV is expected\cite{Behnke:2013lya}.
Therefore, $\Delta m_h \propto E_{cm}^2 / 2 m_{h}$, approximating $E_{Z} \sim E_{CM}/2$ and
the recoil mass resolution gets worse at higher energy.

On the other hand, 
the jet clustering is challenging near $Zh$ threshold due to jet overlap, which limits 
the mass resolution of $Z$ in jet mode.  In this paper, we concentrate on the study 
at $\sqrt{s}=350$ GeV and 500 GeV, where jet are relatively sharp and separated.

We used events generated by Whizard 1.95 with ILC beam parameter\cite{Behnke:2013lya}. 
ILD full detector simulation and reconstruction were used in order to take into account 
signal smearing by detector effects. The underlying low $p_{t}$ hadron background events 
with an average number of events of 0.33 (1.7) at $\sqrt{s}=350 (500)$ GeV
and the beam crossing of 7mrad were taken into account as well.

%%%%%%%%%%%%%%%%%%%%%%%%%%%%%%%%%%%%%%%%%%%%%%%%%%%%%%%%%%%%%%
% Inclusive jet pair study
%%%%%%%%%%%%%%%%%%%%%%%%%%%%%%%%%%%%%%%%%%%%%%%%%%%%%%%%%%%%%%
\section{Recoil mass measurement at $\sqrt{s}=350$ GeV}
For the inclusive jet selection, the $k_t$ jet algorithm implemented in Fastjet\cite{fastjet:242} was employed
with the jet radius of 1.2 and $p_{t,min}=1.0$ GeV/c$^2$ without restricting the number of reconstructed jets.  
Then all combinations of jet pairs were tried to find a jet pair of mass
consistent with $Z$. A good $Z$ jet pair was selected by following conditions; 
(1) Squared transverse momentum($kt_2$) of first jet was between 4000 to 6000 (GeV/c$^2$)$^2$; 
(2) $k_t^2$ of second jet was greater than 500  (GeV/c$^2$)$^2$ ; 
(3) Jet pair energy was between 140 to 180 GeV; 
(4) Corrected mass of jet pair was between 85 to 100 GeV/c$^2$;
(5) No photon with energy greater than 80 GeV in the event.

\begin{figure}[htbp]
  \begin{center}
    \begin{tabular}{c c c}
      \includegraphics[width=0.45\columnwidth]{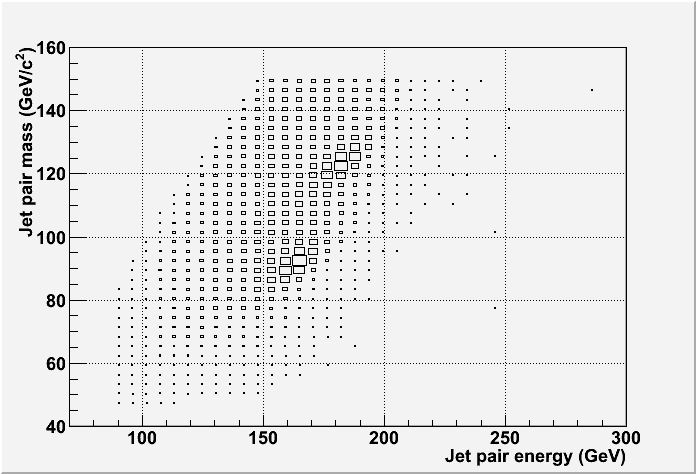} &
      \hspace{0.0001\columnwidth} &
      \includegraphics[width=0.45\columnwidth]{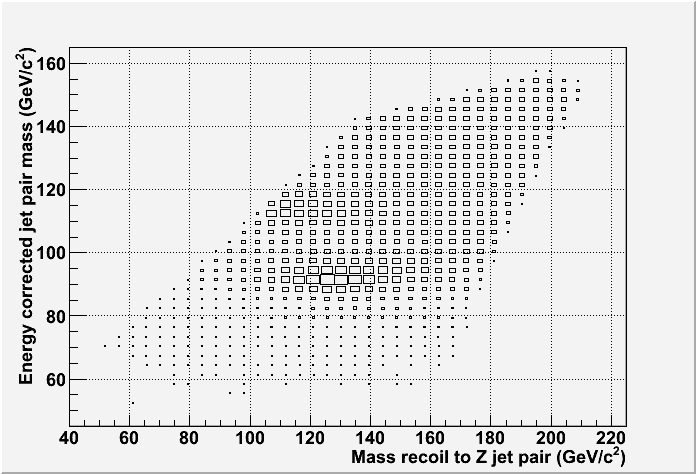} \\
      (a) & & (b)
    \end{tabular}
  \begin{minipage}[c]{0.85\columnwidth}  
  \caption{The scatter plot of the energy and the mass of jet pairs before the energy correction  (a) 
and that for the recoil mass and the mass after the energy correction.  The signals of  
$q\bar{q}h$ events are plotted with the $k_t^2$ cuts described in the text.
\label{figure:jetpair-mass-rm}}
  \end{minipage}
  \end{center}
\end{figure}

With the selection (1) and (2), the clusters of $Z$ and $h$ in the scatter plot of 
the mass and the energy are clearly seen in the case of $q\bar{q}h$ events
as shown in Fig.~\ref{figure:jetpair-mass-rm}a.
In order to remove the observed correlation between the energy and the mass seen in the Fig.~\ref{figure:jetpair-mass-rm}a
the four vector of the jet pair was multiplied by a factor, which was determined to remove 
the linear correlation between the energy and the mass.
The corrected four vector was used to calculate the recoil mass.
The scatter plot of the mass and the recoil mass after the correction is shown in 
Fig.~\ref{figure:jetpair-mass-rm}b.  The recoil mass of $Z$ pair clustered 
near the input Higgs mass of 125 GeV/c$^{2}$, while those for $h$ pair 
were shifted from the right position because the correction factor was determined 
by $Z$ candidate alone, which does not affect the result of this analysis.
The last cut (5) was to remove one of the major background, $e^+e^-\rightarrow q\bar{q}\gamma$.
If more than one $Z$ candidate was found, only first candidate was selected. Note that the output of Fastjet 
is sorted by descending order of jet $p_t$.

\begin{figure}[htbp]
   \begin{center}
   \begin{tabular}{c c c}
      \includegraphics[width=0.45\columnwidth]{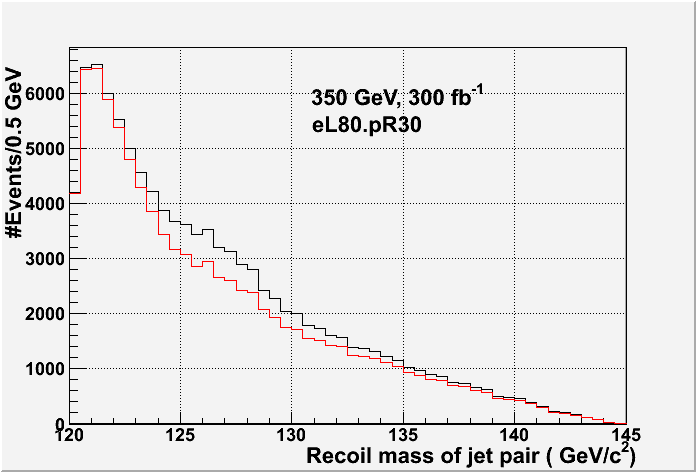} &
      \hspace{0.0001\columnwidth} &
      \includegraphics[width=0.45\columnwidth]{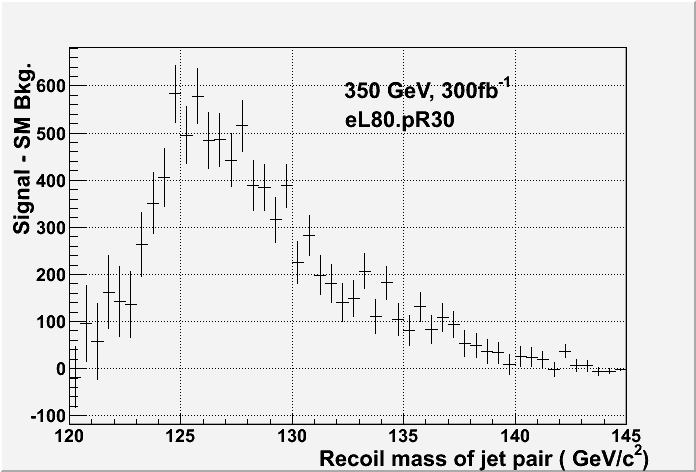} \\
      (a) & & (b)
   \end{tabular}
   \begin{minipage}[c]{0.85\columnwidth} 
   \caption{The left figure, (a), is the recoil mass distribution of the inclusive jet pair at 
$\sqrt{s}=350$ GeV, the integrated luminosity of 300 fb$^{-1}$ and 
the $e^{-}/e^{+}$ beam polarization of -80\%/+30\%. The red histogram is the background only and 
the black histogram is with the signal.  The right figure, (b), is the same histogram as (a), but 
the standard model background being subtracted.
\label{fig:qqh-recoil-mass-350}}
\end{minipage}
\end{center}
\end{figure}

For the background processes, following processes were considered; $e^+e^-\rightarrow 
q\bar{q}$, $q\bar{q}q\bar{q}$, $q\bar{q}\ell\nu$, $\ell\bar{\ell}\nu\bar{\nu}$, $t\bar{t}$, Higgs process other than 
$q\bar{q}h$, and $2f$ and $4f$ processes created by $\gamma\gamma$,  and $3f$ and $5f$ processes created by $e\gamma$ collisions. They were produced for the Snowmass study using the software tools prepared for the ILC DBD\cite{Snowmass:MCSamples2013}.
Fig.~\ref{fig:qqh-recoil-mass-350}a shows the recoil mass distribution of selected events. The red is the 
standard model background and the black is the signal contribution.  Fig.\ref{fig:qqh-recoil-mass-350}b
is the distribution after subtracting the standard model background.  

Finally $q\bar{q}h$ events were selected by requiring the recoil mass between 123 and 133 GeV/c$^2$.
The number of signal and background events selected is summarized in Tab.~\ref{tab:selection350}
signal to noise ratio (S/N) was 0.049 for -80\%(+30\%) $e^{-}(e^{+})$ beam polarization, 
and 0.153 for +80\%(-30\%) $e^{-}(e^{+})$ polarization. Assuming 165 fb$^{-1}$ data taking for 
each beam polarization configuration, the number of signal events was 6194/4169 
for -80\%(+30\%)/+80\%(-30\%) 
beam polarization. $q\bar{q}$, $q\bar{q}q\bar{q}$, and $q\bar{q}\ell\nu$ were major backgrounds 
while  $q\bar{q}q\bar{q}$ contribution dominated in the case of -80\%(+30\%) beam polarization case.
The expected accuracy of the $Zh$ total cross section measurement was 5.9\% and 4.3\% for -80\%/+30\% and 
+80\%/-30\% beam polarization, respectively. Combining two measurement, 3.4\% accuracy was expected 
from the inclusive jet pair measurement.
\begin{table}[t]
\begin{center}
\begin{tabular}{| l | r | r | }
\hline 
process & +80\%/-30\% & -80\%/+30\% \\
\hline
$q\bar{q}h$ & 4169 & 6194 \\
$q\bar{q}$ & 7176 & 18329 \\
$q\bar{q}q\bar{q}$ & 8649 & 59956 \\
$\ell\bar{\ell} q\bar{q}$ & 5921 & 36284 \\
$\ell\bar{\ell}\ell\bar{\ell}$ & 328 & 965 \\
$\ell\bar{\ell}h$ & 722 & 1086 \\
$t\bar{t}$ & 2268 & 4642 \\
$\gamma\gamma/\gamma e^{\pm} \rightarrow 2f/4f$ & 2184 & 4026 \\
\hline
S/$\sqrt{S+N}$ & 23.5 & 17.1 \\
\hline
\end{tabular}
\begin{minipage}[c]{0.8\columnwidth}
\caption{The number of signal and background events after the final selection.
The second and the third row show the number of events with 165 fb$^{-1}$ each for
+80\%(-30\%) and -80\%(+30\%) $e^{-}(e^{+})$ beam polarization.
\label{tab:selection350}}
\end{minipage}
\end{center}
\end{table}

The recoil mass study using $\mu\bar{\mu}$ channel at 350 GeV was reported in ref.\cite{HengneLi:2010} 
using LOI ILD full simulation and the Higgs mass of 120 GeV/c$^{2}$. This study compared the accuracy at 250 GeV and 350 GeV 
and concluded that the expected accuracy of the $Zh$ total cross section ad 250 GeV 
and 350 GeV were similar. As described previously, at 250 GeV, $\Delta\sigma/\sigma=2.6\%$ 
is expected from the recoil mass measurement of $\mu\bar{\mu}$ and $e\bar{e}$ channel. 
Assuming the same accuracy of 2.6\% can be obtained at 350 GeV using $\mu\bar{\mu}$ and $e\bar{e}$ 
channel, we can expect $\Delta\sigma/\sigma=2.1\%$ by combining the result of the recoil mass 
measurement of inclusive two jet. If this measurement is combined with the 2.6\% measurement 
at 250 GeV, we could expect $\Delta\sigma/\sigma=1.6\%$ combining 250 GeV and 350 GeV data taking.

%%%%%%%%%%%%%%%%%%%%%%%%%%%%%%%%%%%%%%%%%%%%%%%%%%%%%%%%%%%%%%
% Inclusive jet pair study : 500 GeV
%%%%%%%%%%%%%%%%%%%%%%%%%%%%%%%%%%%%%%%%%%%%%%%%%%%%%%%%%%%%%%
\section{Recoil mass measurement at $\sqrt{s}=500$ GeV}
At $\sqrt{s}=500$ GeV, the total cross section of $e^{+}e^{-}\rightarrow q\bar{q}h$ process is about 70~fb
when the $e^{-}(e^{+})$ beam polarization is  -80\%(+30\%).  About 35k such events are produced for 500 fb$^{-1}$ 
integrated luminosity.  The resolution of jet energy and recoil mass are not as good as a lower energy 
measurement and leptonic channel, still, we could study this channel thanks to the relatively larger event statistics 
and the monotonic $Z$ and $H$ energy due to $s$-channel 2 body production

The major background processes 
are those by 4-fermion $W^{+}W^{-}$, $Z^0Z^0$, and 2-fermion $q\bar{q}$ processes and it is not easy to 
get good S/N.

\begin{figure}[htbp]
   \begin{center}
   \begin{tabular}{c c c}
      \includegraphics[width=0.45\columnwidth]{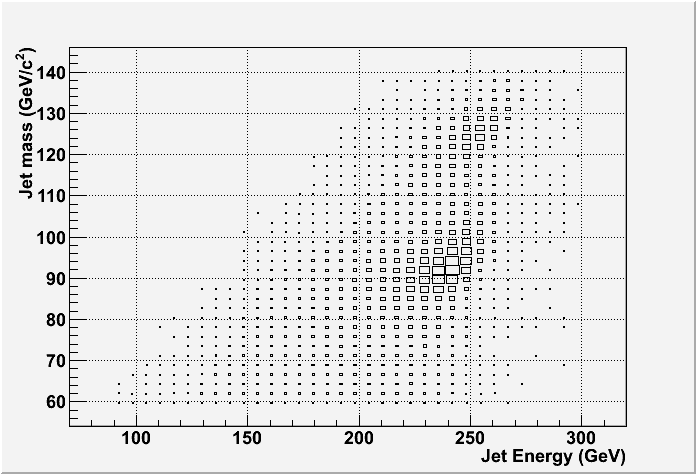} &
      \hspace{0.0001\columnwidth} &
      \includegraphics[width=0.45\columnwidth]{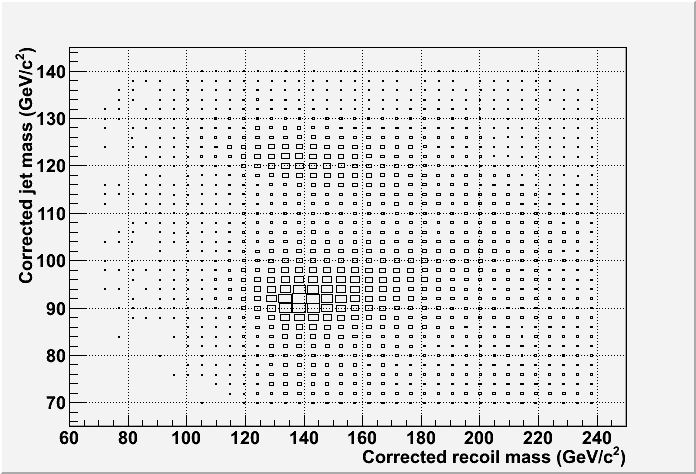} \\
      (a) & & (b)
   \end{tabular}
   \begin{minipage}[c]{0.85\columnwidth} 
  \caption{The scatter plot of the energy and the mass of candidate jets before the energy correction  (a) 
and that for the recoil mass and the mass after the correction.
\label{figure:jet-mass500}}
\end{minipage}
\end{center}
\end{figure}

  The energy of $Z$ from $Zh$ process at this energy is more than 200 GeV and  jets from $Z$ 
are well collimated.  Therefore, we reconstructed hadronically decayed $Z$ as a single jet by 
$k_t$ jet algorithm with a jet radius 1.2. From reconstructed jets, candidate jets were pre-selected 
requiring the jet p$_{t}$ is greater than 50 GeV, the jet mass between 70 and 150 GeV/c$^{2}$ 
and the jet energy between 210 and 300 GeV. The scatter plot of the mass and the energy of 
the selected jet in $q\bar{q}h$ events are shown in Fig.~\ref{figure:jet-mass500}a.

With a fixed jet radius, both jet mass and jet energy were reduced if particles from $Z$ escaped 
from the jet radius, thus a positive correlation between mass and energy was seen as shown in Fig.~\ref{figure:jet-mass500}.
This correlation was removed by scaling jet momenta with a factor which linearly depended on jet energy.
After the correction, a better separation between $Z$ jet and non-$Z$ jet were 
achieved as seen in Fig.~\ref{figure:jet-mass500}b.  

For the final selection, 
we further required (1) the corrected jet mass between 87 and 105 GeV/c$^{2}$, (2) the maximum 
energy of $\gamma$ in the event is less than 100 GeV ( to suppress $q\bar{q}\gamma$ background events), 
(3) Number of particles in the jet is greater than 20, 
(4) Jet angle satisfies $|\cos\theta_{jet}|<0.7$.  

\begin{figure}[htbp]
   \begin{center}
   \begin{tabular}{c c c}
      \includegraphics[width=0.45\columnwidth]{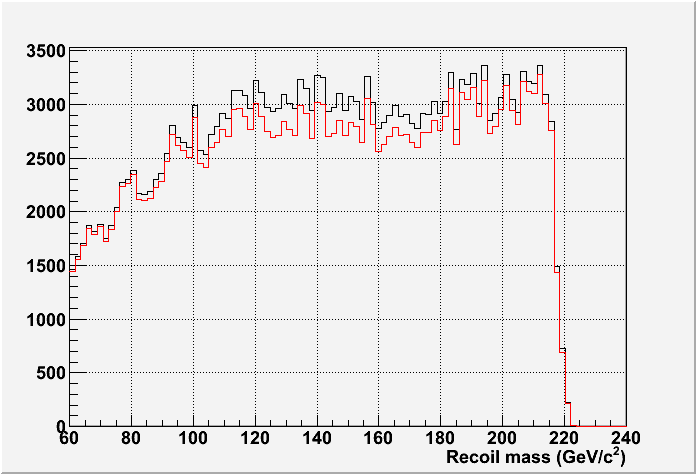} &
      \hspace{0.0001\columnwidth} &
      \includegraphics[width=0.45\columnwidth]{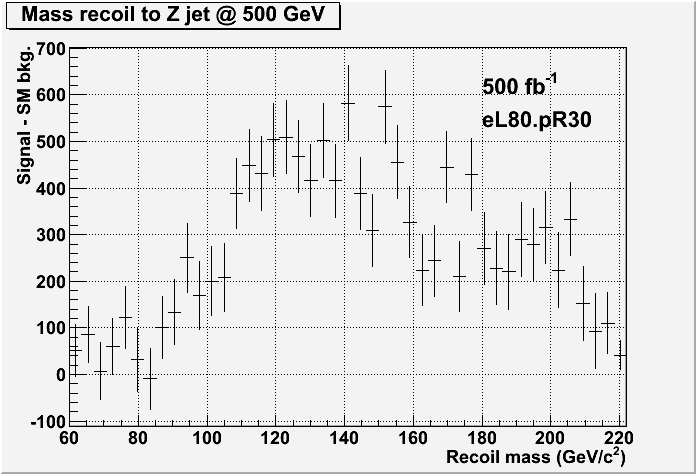} \\
      (a) & & (b)
   \end{tabular}
   \begin{minipage}[c]{0.85\columnwidth} 
   \caption{The left figure is the recoil mass distribution of selected events at $\sqrt{s}=500$ GeV 
with $e^{-}(e^{+})$ beam polarization of -80\%(+30\%) and 500 fb$^{-1}$ integrated luminosity.  The red histogram 
is for the standard model processes and the black histogram is with the $q\bar{q}h$ events added.
The right figure is the distribution after subtracting the background contribution. Note that the jet momenta
before the correction is used for the recoil mass calculation in this figure.
\label{fig:incjet-recoil-mass-500}}
\end{minipage}
\end{center}
\end{figure}

The recoil mass distribution of selected events without/with background subtraction are shown in Fig.~\ref{fig:incjet-recoil-mass-500}. In this figure, the jet momenta before the correction was used to calculate the recoil mass.
As background processes, the standard model samples produced for the 
ILC DBD study\cite{Behnke:2013lya} were considered. They included $2f/4f/6f$ standard model processes 
$4f/5f$ final states produced by $\gamma\gamma$ and $\gamma e^{\pm}$ collisions, 
and $f\bar{f}h$ process except $q\bar{q}h$.

\begin{figure}[htbp]
   \begin{center}
   \begin{tabular}{c c c}
      \includegraphics[width=0.45\columnwidth]{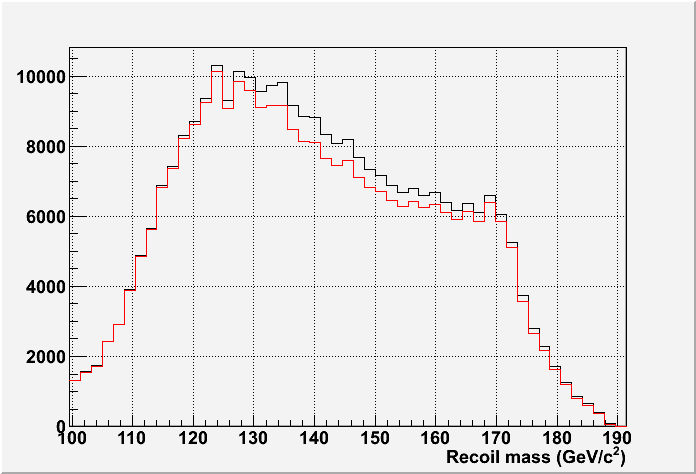} &
      \hspace{0.0001\columnwidth} &
      \includegraphics[width=0.45\columnwidth]{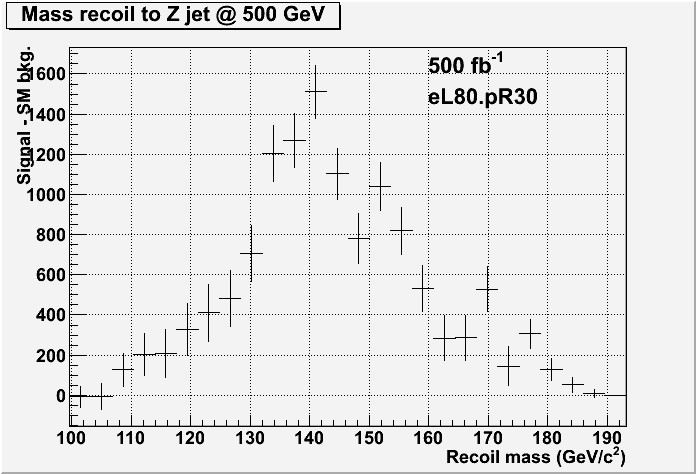} \\
      (a) & & (b)
   \end{tabular}
   \begin{minipage}[c]{0.85\columnwidth} 
   \caption{Same as Fig.~\ref{fig:incjet-recoil-mass-500}, but the corrected jet momenta were used 
for the recoil mass calculation.
\label{fig:incjet-recoil-mass-500-withMassCorrec}}
\end{minipage}
\end{center}
\end{figure}

As the final selection, events with the recoil mass between 100 and 210 GeV/c$^{2}$ were selected. 
The S/N of this selection was 11113/175437=0.063.  43\% backgrounds were due to 4-quark events through $ZZ$ and $WW$ processes. Other 4-fermion 
processes and 2-fermion hadron events constitutes 26\% and 27\% of background events, respectively.  
The number of events selected are summarized in Table~\ref{tab:500selection}.
The signal significance for 500 fb$^{-1}$ is 3.9\%.
The $e^{+}(e^{-})$ beam polarization opposite to this analysis, namely +80\%(-30\%), 
could reduce some of the $4f$ background which comes from $W^{+}W^{-}$ process, 
still there remains significant background events due to $ZZ\rightarrow 4q$ and $2q$ processes. 

The distribution of the recoil mass calculated from the corrected jet momenta is shown in Fig.~\ref{fig:incjet-recoil-mass-500-withMassCorrec}.  The S/N is similar to the case of Fig.~\ref{fig:incjet-recoil-mass-500}. From the events 
with the mass between 130 to 170 GeV/c$^{2}$, we obtained the signal significance of 3.9\% for 
500 fb$^{-1}$ with $-80\%(+30\%)$ $e^{-}(e^{+})$ beam polarization.

\begin{table}[t]
\begin{center}
\begin{tabular}{| l | r | }
\hline 
process & No. of events  \\
\hline
$q\bar{q}h$ & 11113  \\
$f\bar{f}h$ & 338 \\
$q\bar{q}$  & 47377 \\
$q\bar{q}q\bar{q}$ & 121086 \\
$q\bar{q}q\bar{q}q\bar{q}$ & 6357 \\
$\gamma\gamma/\gamma e^{\pm} \rightarrow 2f/4f$ & 277 \\
\hline
S/$\sqrt{S+N}$ & 25.7 \\
\hline
\end{tabular}
\caption{The number of signal and background events after selection.\label{tab:500selection}}
\end{center}
\end{table}

% =========================================================================
\section{Conclusion}
In this paper, a feasibility to measure the total cross section of the Higgs-strahlung process using 
the hadronic decay mode of $Z$ was studied.  At $\sqrt{s}=350$ GeV, Higgs peak in jet recoil mass 
distribution could be seen; Combining 165 fb$^{-1}$ measurements with the $e^{+}(e^{-})$ beam polarization
of -80\%(+30\%) and +80\%(-30\%),respectively, $\Delta\sigma_{Zh}/\sigma_{Zh} = 3.4\%$
was expected.  At 500 GeV, it was hard to see a clear Higgs peak in the jet recoil mass 
distribution. However, from the excess of events in $q\bar{q}h$ like events, 
$\Delta\sigma_{Zh}/\sigma_{Zh} = 3.9\%$ was expected for 500 fb$^{-1}$ measurements 
with $e^-(e^+)$ beam polarization of -80\%(+30\%).
The analysis was based on a cut base event selection.  Further improvement would be possible 
with more sophisticated analysis.  

%At the energy around $Zh$ production threshold, 
%separation of $Z$ jets and $h$ jets are challenging 

% =========================================================================
\section*{Acknowledgments}
The author would like to thank Keisuke Fujii, Timothy Barklow, Junping Tian and Taikan Suehara 
for encouragement and discussion for this analysis, and ILD group where this study is based on.
This work was supported in part by the Japan Society for Promotion of Science (JSPS) Grant-in-Aid
for Specially Promoted Research No. 23000002.

% =========================================================================
\bibliographystyle{utphys}
\bibliography{qqhmain}

\end{document}